\documentclass[aps,fleqn,amsmath,amssymb]{revtex4}

\newcommand{\bed}{\[}
\newcommand{\eed}{\]}
\newcommand{\beq}{\begin{equation}}
\newcommand{\eeq}{\end{equation}}
\newcommand{\beqa}{\begin{eqnarray}}
\newcommand{\eeqa}{\end{eqnarray}}
\newcommand{\ket} [1] {\vert #1 \rangle}
\newcommand{\bra} [1] {\langle #1 \vert}
\newcommand{\braket}[2]{\langle #1 | #2 \rangle}
\newcommand{\proj}[1]{\vert{#1} \rangle \langle {#1} \vert}
\newcommand{\mean}[1]{\langle #1 \rangle}
\newcommand{\gras}[1]{\mathbf{#1}}

\newcommand{\tr}{\mathop{\mathrm{tr}}}

\usepackage{graphicx}
\usepackage[mathscr]{eucal}
\usepackage{amsmath}
\usepackage{amssymb}
\usepackage{epstopdf}
\usepackage{epsfig}
\begin{document}

\title{Quantum Relative States}
\author{N. Gisin}
\author{S. Iblisdir}
\affiliation{ GAP-Optique, Universit\'e de Gen\`eve, 1211 Gen\`eve, Switzerland}
\email{nicolas.gisin@physics.unige.ch,sofyan.iblisdir@physics.unige.ch}

\date{\today}

\begin{abstract}

We study quantum state estimation problems where the reference
system with respect to which the state is measured should itself
be treated quantum mechanically. In this situation, the difference
between the system and the reference tends to fade. We investigate
how the overlap between two pure quantum states can be optimally
estimated, in several scenarios, and we re-visit homodyne
detection.
\end{abstract}

\maketitle

\section{Introduction}
Today, entanglement is widely recognized as one of the
characteristic feature, if not the essence, of quantum physics
\cite{Terhal,nielsen}. Historically, the experimental paradigm of
entanglement has long been Bell tests \cite{bell,belltests}, that
is, experiments that demonstrate quantum non-locality. A more
modern example is quantum teleportation \cite{teleportation}. The
beauty of teleportation is that a quantum state dissolves at some
point of space-time and appears at another without ever existing
in-between. Indeed, neither the quantum system held by the
receiver, nor the classical information, that make teleportation
possible, contain any information about the original state.

A fundamental problem tackled by quantum information Science is to
characterize entangled \emph{states}, and much has been learnt
along this line of research \cite{nielsen}. However, entanglement
is also a feature of some measurements, that we will refer to as
coherent measurements, characterized by self adjoint operators
whose eigenvectors are entangled states. Interestingly, both
aspects of entanglement, i.e. states exhibiting quantum
correlations and coherent measurements, are exploited in an
essential way in quantum teleportation\footnote{Despite all the
limitations of intuition in quantum physics, let us elaborate on
this. Roughly, entanglement provides the parties with correlations
strong enough for them to be able to always give the same answer
whenever asked the same question (the typical singlet state always
provides opposite measurement results, whenever measurement along
parallel axes are performed). Now, the coherent Bell measurement,
used in quantum teleportation, does something like asking "{\it
how similar they you?}" to one of the entangled system and to the
system to be teleported. If the answer happens to be: {\it we are
alike}, i.e. would we receive the same question, we would give the
same answer, then, clearly, Bob's system would response to any
measurement in the same way the original system would, hence
achieving teleportation. Next, if the answer is: we are alike up
to a standard symmetry, then teleportation succeeds as soon as the
receiver, Bob, gets the information about which standard symmetry.
This illustrates how quantum teleportation exploits the dual
aspect of entanglement, i.e. that both aspects of entanglement are
equally essential although one aspect received much more attention
than the other, because of the historical role played by Bell
inequalities. }

A property of coherent measurements, that we will be interested in
here, is that they allow to measure \emph{relative} properties of
a set of quantum systems without gaining information about the
individual subsystems. In contrast, there is no non-trivial manner
to measure a relative property of classical systems without
actually measuring each system and computing the relative property
from the measurement outcomes. For example, given two classical
arrows, there is no way to find out the angle between the arrows
without gaining information about the direction of each arrow (at
least in principle, in practice one can always forget about
classical information). What happens, in this classical setting,
is that the first direction is macroscopic enough that it serves
as a reference axis, with respect to which the direction of the
second arrow is measured. This measurement can be performed with
very high precision since the second arrow is macroscopic enough
that it can be considered classical.

Every measurement on a quantum system can be thought of as a
measurement of some property of this system with respect to a
reference system. This reference system is usually that
macroscopic that one can safely disregard its quantum properties.
We here want to consider the case where \emph{both} the reference
system and the measured system are treated quantum mechanically.
In a sense, we want to ``quantise'' the reference system of a
quantum measurement.

We will treat two classes of relative state estimation problems:
the quantum analogue of the estimation of the angle between two
arrows, and homodyne measurements. In Section \ref{sec:relmeas},
we will address the issue of optimally estimate the (modulus of
the) scalar product between two qubit states, extending on the
work of Ref. \cite{bart04}. We will consider (i) the situation
where each qubit state is represented by identically prepared
qubits, (ii) the situation where one state is represented by an
orthogonally prepared qubit pair, (iii) we will discuss how the
problem can be generalised to qudits. In Section
\ref{sec:homodyne}, we will discuss why homodyne measurements can
be thought of as relative state measurements. We conclude in
Section \ref{sec:conclusion}.

\section{Relative state measurements}\label{sec:relmeas}

\subsection{Each state is represented by identically prepared qubits}

Consider the problem of estimating the angle between two
directions \cite{bart04}. The first direction is represented by
$N$ qubits identically prepared in some state $\ket{\psi_1}$ and
the second direction is represented by $M$ qubits identically
prepared in some state $\ket{\psi_2}$. We will assume $M \geq N$.
Our aim will be to estimate at best  the value of
$|\braket{\psi_1}{\psi_2}|^2$. This problem can be thought of as
the estimation of the state of a quantum system. e.g. the first
one, relative to an axis which is it self quantum, the second
system.

We will construct a positive operator valued measure $\{P(x)\}$:
\bed P(x) \in \mathscr{B}(\mathscr{H}^{\otimes N+M}), \; P(x) \geq
0; \; \int_0^1 dx \; P(x)= \gras{1}^{\otimes N+M}, \eed where
$\mathscr{H}$ denotes the Hilbert space of a qubit, and
$\mathscr{B}(\mathscr{H}^{\otimes N+M})$ denotes the space of
(bounded) operators acting on $\mathscr{H}^{\otimes N+M}$.
$\gras{1}$ denotes the $2 \times 2$ identity matrix.

When the outcome $P(x)$ comes out, the value $x$ is guessed. As a
figure of merit, we will take the mean variance:

\beq \Delta({P(x)})=\int dx \; d\psi_1 \; d\psi_2 \;
\textrm{Prob}(x \mid \psi_1, \psi_2) \;
(x-\mid\braket{\psi_1}{\psi_2}\mid^2)^2 \eeq

$\textrm{Prob}(x \mid \psi_1, \psi_2)$ denotes the conditional
probability to get the outcome $x$ when a measurement is performed
on qubits prepared in the states $\psi_1$, $\psi_2$. More
technically, this variance can be re-expressed as:

\beq \Delta({P(x)})=\int dx \; dg_1 \; dg_2 \;
\bra{\psi_0^{\otimes N+M}} \pi_N^+(g_1)^* \otimes \pi_M^+(g_2)^*
P(x)
 \pi_N^+(g_1) \otimes \pi_M^+(g_2)
\ket{\psi_0^{\otimes N+M}} (x-|\bra{\psi_0} \pi(g_1)^* \pi(g_2)
\ket{\psi_0}|^2)^2. \eeq In this expression, $g_1$ and $g_2$
represent SU(2) elements and $dg_1, dg_2$ represent  the Haar
measure over $SU(2)$. $\pi$ denotes the natural representation and
$\pi_{N}^+$ denotes the irreducible representation obtained by
restriction of $\pi^{\otimes N}$ onto the symmetric subspace of
the space of $N$ qubits, $\mathcal{H}_N^+$. Actually $\pi_{N}^+$
is the spin-$j$ irreducible representation, with $N=2j$.
$\ket{\psi_0}$ is some fiducial state.

From any povm $P(x)$, one can construct another povm:
\beq
Q(x)=\int dg \; (\pi^{\otimes N} (g)^* \otimes \pi^{\otimes M} (g)^*) P(x)
(\pi^{\otimes N} (g) \otimes \pi^{\otimes M} (g)).
\eeq
that achieves the same error variance as $P(x)$.
Clearly, $[Q(x),\pi_N^+(g) \otimes \pi_M^+(g)]=0 \hspace{0.5cm} \forall x,g$.
Using the SU(2) Clebsch-Gordan series, the latter commutation relation can be rewritten as
\beq\label{eq:cova}
[Q(x), \bigoplus_{k=M-N}^{M+N} \pi_k^+(g)]=0 \hspace{1cm} \forall x,g.
\eeq
A nice property of the second argument of the commutator (\ref{eq:cova})
is that no representation appears more than once. Consequently (Shur's lemma),
$Q(x)$ has the following \emph{diagonal} form
\beq\label{eq:structure}
Q(x)= \sum_{k=M-N}^{M+N} q_k(x) \gras{1}_k,
\eeq
where $\gras{1}_k$ is the projector onto the irreducible subspace supporting $\pi_k^+$.

The condition that $\{ Q(x)\}$ should be a povm is then expressed as
\beqa
&& \int_0^1 dx \; q_k(x)=1, \hspace{1cm} \forall k=M-N, \ldots, M+N,\\
&& q_j(x) \geq 0 \; \forall x.
\eeqa

The score $\Delta(\{Q(x)\})$ can now be written as \beq
\Delta(\{Q(x)\})=\sum_{k=M-N}^{M+N} \int_0^1 dx \; q_k(x) P_k(x),
\eeq where $P_k(x)$ is a degree-$2$ polynomial in $x$: $P_k(x)=
I_k^0 x^2 - 2 I_k^1 x +I_k^2$. Explicit expressions for the
quantities $I_k^{\alpha}$ are given in Appendix \ref{sec:ikj}.

Clearly, the optimal povm is given by
$q_k(x)=\delta(x-x_k^{\textrm{min}})$, where
$P_k(x_k^{\textrm{min}})=\textrm{min}_{0 \leq x \leq 1} P_k(x)$.
Since the $P_k$ are polynomials of degree 2, the
$x_k^{\textrm{min}}$ are readily calculated and one finds that
\beq\label{eq:varformula}
\Delta^{\textrm{opt}}(N,M)=\sum_{k=M-N}^{M+N}
(I_k^2-\frac{(I_k^1)^2}{I_k^0}). \eeq Values
of$\Delta^{\textrm{opt}}(N,M)$ for some values of $N,M$ are given
in Table \ref{tab:para}.

\begin{table}[h]
\begin{center}
\begin{tabular}{|c||c|c|c|c|c|c|c|c|}
\hline
$N$
& 1
& 1
& 1
& 2
& 2
& 7
& 20
& 1\\
\hline
$M$
& 1
& 2
& 300
& 2
& 3
& 7
& 20
& $\infty$\\
\hline
$\Delta^{\textrm{opt}}(N,M) \times 10^{2}$
& 7.41
& 6.94
& 5.57
& 6.25
& 5.83
& 3.29
& 1.45
& 5.56\\
\hline
\end{tabular}
\end{center}
\caption{Minimal variance $\Delta(N,M)$.} \label{tab:para}
\end{table}

Let us now comment a bit on Eq.(\ref{eq:structure}). This equation
tells us that the best strategy is to use a measurement whose
elements are projectors onto subspaces invariant under
$\pi^{\otimes N+M}$. Clearly, the space of the $N+M$ particles at
hand also supports a representation of the permutation group of
$N+M$ objects, $\textrm{Sym}(N+M)$. Now it is an important result
of representation theory that the algebra linearly generated by
all unitaries $\pi^{\otimes N+M}(g)$ and the algebra generated by
permutation operators on $\mathcal{H}^{\otimes N+M}$ are commutant
of each other. Consequently, they have \emph{common} invariant
subspaces \cite{zhelobenko}. In our case, this means that the
elements of our povm project onto subspaces that are invariant
under permutation of particles. We interpret this fact as follows.
No preferred reference frame is available to estimate the angle
between two directions, but it is known which particle belongs to
the set indicating the first (resp. the second) direction.
Therefore, it seems natural that the only kind of properties that
can be measured are those related the permutations that can be
carried on the particles.

Our point is more easily illustrated in the case where $N=1,M=1$.
The optimal povm is then made of two pieces, the singlet, a state
which changes sign when a permutation is applied, and the triplet,
which remains unchanged when the a permutation is applied. Thus,
our measurement actually tests permutation properties of our system,
on the basis of which a guess of the relative angle is made: the
singlet representation of $\textrm{Sym}(2)$ makes us guess that
$|\braket{\psi_1}{\psi_2}|^2=1/3$ (the states are rather antiparallel),
and the triplet representation makes us guess that $|\braket{\psi_1}{\psi_2}|^2=5/9$
(the states are rather parallel). 


\subsection{One state is represented by one qubit, the other by two orthogonally prepared qubits}

We now turn to the situation where one direction is specified by two anti-parallel qubits.
Thus, let $\{ \psi_0, \psi_1 \}$ denote an orthonormal basis of $\mathscr{H}$, the
Hilbert space of one qubit. One direction is specified by an element of SU(2), $g_1$ say,
and the other direction is specified by $g_2 \in$  SU(2). We are now given the state
$\pi(g_1)^{\otimes 2} \ket{\psi_0,\psi_1} \pi(g_2) \ket{\psi_0}$, and we want (again)
to estimate at best $\bra{\psi_0} \pi(g_1)^* \pi(g_2) \ket{\psi_0}$. Again, we are looking
for a povm $\{P(x) \in \mathscr{B}(\mathscr{H}), P(x) \geq 0, \int_0^1 dx P(x)=\gras{1}^{\otimes 3} \}$.
The figure of merit has a form similar to the one we had before:

\beq
\Delta= \int_0^1 dx \int dg_1 \int dg_2
\bra{\psi_0,\psi_1,\psi_0}
\pi(g_1)^{\otimes 2} \otimes \pi(g_2)
P(x)
\pi(g_1)^{\otimes 2^*} \otimes \pi(g_2^*)
\ket{\psi_0,\psi_1,\psi_0}
(x-|\bra{\psi_0} \pi(g_1)^* \pi(g_2) \ket{\psi_0} |^2)^2.
 \eeq

Again we can restrict to covariant measurement and assume that
\beq [P(x), \pi^{\otimes 3}(g)]=0, \hspace{1cm} \forall x,g. \eeq
The details of the extremisation can be found in appendix
\ref{sec:antipara}. The main result is that, perhaps surprisingly,
we find essentially the same mean variance as in the case where each
state is represented by two identically prepared qubits \footnote{This result is consistent with what has been found in \cite{bart04},\cite{lind05}. Actually, careful calculations indicate tiny differences between the anti-parallel case and the parallel case. The antiparallel case exhibits a slightly lower variance (a difference emerges from the tenth digit). In contrast, if one considers the fidelity as a figure of merit as in \cite{baga05}, then parallel pairs are slightly better. We have not investigated these differences further.} Unfortunately, we don't have any intuition on why parallel and antiparallel pairs should perform as well or not  for our problem. More generally, the differences between antiparallel qubit pairs and parallel qubit pairs in quantum estimation theory are still poorly understood \cite{brau04}.

\subsection{Generalisation to qudits}

The foregoing analysis can be straightforwardly extended to qudit
systems of arbitrary finite dimension $d$. In appendix
\ref{sec:calculqudits}, we have computed the (generalisation of
the) formula (\ref{eq:varformula}) in the case that $N=M=1$. We
have found \beq\label{eq:meanvard}
\Delta=\frac{2}{d^2+d}-\frac{(d-2)^2}{2d(d+1)^2(d-1)}-\frac{(d+3)^2}{2d(d+1)^3}.
\eeq We see that the mean variance decreases with $d$ as $\approx
1/d^2$. The fact that this variance should decrease with $d$ could
be expected because when the dimension increases, the overlap
between two randomly drawn states tends (on average) to $0$, i.e.
the states are increasingly orthogonal, and thus easier to
estimate. We also note that the povm consists again on projectors
onto subspaces invariant under permutations of particles, i.e. the
overlap between two quantum states is estimated upon testing
permutation properties.

\section{Homodyne Detection}\label{sec:homodyne}

We now wish to describe how homodyne detection can be thought of
as a relative state measurement. (See also Ref.
\cite{ahar67}\cite{molm97}) In a homodyne measurement
\cite{wall94}, two e-m fields impinge on the two input ports of a
balanced beam splitter. One of the input is generally referred to
as ``signal'', and the other as ``reference''. We will assume that
the signal and the reference have the same frequency and the same
polarisation. A photodetector is placed at each output port of the
beam splitter. This scheme aims at measuring a quadrature of the
signal field from the difference of the two photocurrents read on
the detectors. Let $a,b$ denote the annihilation operators for the
two input ports, and $c,d$ the annihilation operators for the
output ports. The observable that is actually measured by the
homodyne setup is \beq\label{eq:obshomod} c^*c-d^*d=a^* b+ b^* a,
\eeq where $c=\frac{1}{\sqrt{2}}(a+b),d=\frac{1}{\sqrt{2}}(a-b)$.

The reference field is assumed to be in a coherent state
$\ket{\psi_r}=\ket{\beta e^{i \theta}}$, where $\beta$ and $\theta$ are
two known real numbers. It is also assumed that $\beta$ is so large that
$b \approx \mean{b}=\beta e^{i \theta}$, i.e. the reference is a
\emph{classical} field. Then the homodyne setup measures the observable
$\beta (a^* e^{i \theta}+a e^{-i \theta})$, and thus indeed corresponds
to measuring a quadrature of the signal field. We can choose the
quadrature we wish to measure upon tuning the phase $\theta$.

Let $\ket{\psi_s}=\sum_n \psi_n \frac{a^{*n}}{\sqrt{n!}}
\ket{\textrm{vac}}$, denote the state of the signal. It is a
remarkable fact that the probability to get a given outcome, say
$K$, is invariant under the transformation $a \to e^{i \phi} a , b
\to e^{i \phi} b$, for arbitrary values of $\phi$, or equivalently
\bed \ket{\psi_r}= \ket{\beta e^{i \theta}} \to
\ket{\psi_r(\phi)}= \ket{\beta e^{i (\theta+\phi)}}, \hspace{1cm}
\ket{\psi_s}=\sum_n \psi_n \frac{a^{*n}}{\sqrt{n!}}
\ket{\textrm{vac}} \to \ket{\psi_s(\phi)}=\sum_n \psi_n
\frac{a^{*n} e^{i n\phi}}{\sqrt{n!}} \ket{\textrm{vac}}. \eed

Thus, whatever convention we choose for the absolute phase of the
e-m field, this convention does \emph{not} affect in any manner
the consistency of homodyne measurement. For example, we could
choose the convention where the field is described by

\beq\label{eq:ranphi} \int_0^{2 \pi} \frac{d\phi}{2 \pi}
\proj{\psi_r(\phi)} \otimes \proj{\psi_s(\phi)}. \eeq

We will restrict the remaining of the discussion to the case where
the state we want to measure is a coherent state that we will denote
$\ket{\alpha}$. Then, one can re-write the state (\ref{eq:ranphi}) in
number state basis as
\beq
\sum_{k=0}^{\infty}\frac{e^{-(|\alpha|^2+|\beta|^2)} \sqrt{|\alpha|^2+|\beta|^2}}{k!}  \ket{k}_c \bra{k},
\eeq
where $\ket{k}_c=\frac{(\alpha a^*+ \beta b^*)^k}{\sqrt{|\alpha|^2+|\beta|^2} \sqrt{k!}}
\ket{\textrm{vac}}$ denotes a state of $k$ photons in the mode
$\frac{\alpha a^*+ \beta b^*}{\sqrt{|\alpha|^2+|\beta|^2}}$.



Assume that the mean photon number $|\alpha|$ is known,
$\textrm{arg}(\alpha)$ is the quantity (phase) we like to measure.
But with respect to what? To a reference $\ket{\beta}$. In words,
instead of thinking of a signal and a reference system, we can
equivalently think of a Poisson distribution of qubits all in the
state $\ket{\psi} \propto \alpha \ket{0}+ \beta \ket{1}.$ In this
description, the difference between the reference and the signal
has completely disappeared. If the mean number of photons $\mid
\beta \mid^2$ is known and very large, then homodyne measurement
turns to be an estimation problem for qubits on a circle of the
Bloch sphere \cite{brau04}.

\section{Conclusion}\label{sec:conclusion}

In summary, we have considered relative state estimation problems,
where the reference system is itself quantum. We emphasized how
general this concept of relative state is and that it conveys an
aspect of entanglement dual to the most studied quantum
correlation between subsystems. More specifically, we have
investigated the problem of estimating the overlap between two
(pure) quantum states in various scenarii.

In the case where each state is represented by identically
prepared qubits, we have noticed a connection between  optimal
strategies and measurements testing permutation properties of the
systems at hand. It would be interesting to investigate this
connection further in other estimation problem.

We have also seen antiparallel qubit pairs and parallel qubit
pairs play are equivalent when used as a reference axis with
respect to which a qubit is measured. It is an interesting open
problem to provide a qualitative explanation for this fact.

We have also revisited homodyne measurements, and discuss why it
is a relative state measurement.

\section{Acknowledgements}

We thank E. Bagan, R. Munoz-Tapia for interesting discussions.
During completion of this work, N. Lindner and coworkers have
independently obtained similar results \cite{lind05}. We thank
them for fruitful discussions and correspondence. Financial
support from the European project RESQ and the swiss NCCR "Quantum
Photonics" are gratefully acknowledged.

\appendix

\section{Evaluations of $I_k^i$}\label{sec:ikj}

Let us start with $I_k^0$. From Schur's lemma, we find that \beqa
I_k^0 &=&  \frac{1}{\textrm{dim}\mathcal{H}_M^+}
\tr (\gras{1}_k (\proj{\psi_0^{\otimes N}} \otimes \gras{1}_M)).\\
 &=& \frac{1}{\textrm{dim}\mathcal{H}_M^+} \int_{SU(2)} dg
 \tr (\gras{1}_k (\pi^+_N(g) \proj{\psi_0}^{\otimes N} \pi^+_N(g)^* \otimes \gras{1}_M))\\
 &=& \frac{\tr \gras{1}_k}{\textrm{dim}\mathcal{H}_M^+
 \textrm{dim}\mathcal{H}_N^+}.
\eeqa
 Similarly, one shows that
\beq\label{eq:ik1} I_k^1 =
\frac{1}{\textrm{dim}\mathcal{H}^+_{M+1}} \tr( (\gras{1}_k \otimes
\proj{\psi_0}) ( \proj{\psi_0^{\otimes N}} \otimes
\gras{1}_{M+1})),
 \eeq
\beq\label{eq:ik2}
 I_k^2 =
\frac{1}{\textrm{dim}\mathcal{H}^+_{M+2}} \tr( (\gras{1}_k \otimes
\proj{\psi_0^{\otimes 2 }})( \proj{\psi_0^{\otimes N}} \otimes
\gras{1}_{M+2})).
\eeq

Straightforwardly, $I_k^0=\frac{k+1}{(N+1)(M+1)}$. Unfortunately,
we were not able to find expressions as simple for $I_j^1$ and
$I_j^2$. However, a direct computation shows that
\beq\label{eq:ik1cg} I_j^1= \frac{1}{M+2} \sum_{m=-j}^{+j}
|C_{(N/2,N/2)(M/2,m-N/2)}^{(j,m)}|^2
|C_{(M/2,m-N/2)(1/2,1/2)}^{((M+1)/2,m-N/2+1/2)}|^2, \eeq

\beq\label{eq:ik2cg} I_j^2= \frac{1}{M+3} \sum_{m=-j}^{+j}
|C_{(N/2,N/2)(M/2,m-N/2)}^{(j,m)}|^2
|C_{(M/2,m-N/2)(1,1)}^{((M+2)/2,m-N/2+1)}|^2, \eeq where
$C_{(j_1,m_1)(j_2,m_2)}^{j,m}=\braket{j,m}{j_1,m_1;j_2,m_2}$
denote Clebsch-Gordan coefficients.

\section{Generalisation to qudits}\label{sec:calculqudits}

The irreducible representations of SU(d) are labelled by $d$-uples
of positive integers $m_1,\ldots,m_d$ satisfying $m_1 \geq \ldots
\geq m_d$ \cite{zhelobenko}. These $d$-uples are called the
highest weights of the representations. We can moreover always
choose $m_d=0$. The Clebsch-Gordan series for $\pi_N^+ \otimes
\pi_M^+$ now reads \beq \pi_N^+ \otimes \pi_M^+ \approx
\bigoplus_{0 \leq k \leq \textrm{min}\{M,N\}}
\pi(M+N-k,k,0,\ldots,0). \eeq Again, no representation appears
more than once in this series, so that $Q(x)$ assumes again a
diagonal form.

The relations (\ref{eq:ik1})-(\ref{eq:ik2}) still hold. But giving
the analogue of Eqs.\;(\ref{eq:ik1cg})-(\ref{eq:ik2cg}) involves
dealing with SU($d$) Clebsch-Gordan coefficients for $d
>2$, which is a heavy business. Therefore we didn't carry our analysis as
far as for the qubit case. There are however some interesting
situations where the expressions (\ref{eq:ik1})-(\ref{eq:ik2}) can
be calculated relatively easily, such as the case where $N=M=1$,
which we will discuss now. First we need expressions for
$\gras{1}(1,1,0,\ldots,0)$, the projector onto the antisymmetric
subspace of two qudits, $\gras{1}(2,0,0,\ldots,0)$, the projector
onto the symmetric subspace of two qudits, and
$\gras{1}(3,0,0,\ldots,0)$, the projector onto the symmetric
subspace of three qudits. We have: \beqa \gras{1}(1,1,0,\ldots,0)
&=& \frac{1}{2} \sum_{k,l=1}^d
(\ket{kl}-\ket{lk})\bra{kl},\\
\gras{1}(2,0,\ldots,0) &=&
\frac{1}{2}
\sum_{k,l=1}^d
(\ket{kl}+\ket{lk})\bra{kl},\\
\gras{1}(3,0,\ldots,0) &=&
\frac{1}{6}
\sum_{k,l,m=1}^d
(\ket{klm}+
\ket{kml}+
\ket{lmk}+  \ket{lkm}+
\ket{mkl}+
\ket{mlk}
)\bra{klm}.\\
\eeqa Then, using the fact that
$\textrm{dim}\mathcal{H}^+_{N}=(d-N+1)!/N!(d-1)!$
\cite{zhelobenko}, we find \beqa I^0(1,1,0,\ldots,0) &=&
\frac{1}{(\textrm{dim}\mathcal{H})^2}\tr \gras{1}(1,1,0,\ldots,0)=
\frac{d-1}{2d},\\
I^0(2,0,\ldots,0) &=&
\frac{1}{(\textrm{dim}\mathcal{H})^2}\tr \gras{1}(1,1,0,\ldots,0)=
\frac{d+1}{2d},\\
I^1(1,1,0,\ldots,0) &=&
 \frac{1}{\textrm{dim}\mathcal{H}^+_{2}} \tr( (\gras{1}(1,1,0,\ldots,0) \otimes \proj{\psi_0})( \proj{\psi_0} \otimes \gras{1}(2,0,\ldots,0)))
 = \frac{d-2}{2d(d+1)},\\
I^1(2,0,\ldots,0) &=&
 \frac{1}{\textrm{dim}\mathcal{H}^+_{2}}
 \tr( (\gras{1}(2,0,\ldots,0) \otimes \proj{\psi_0})( \proj{\psi_0} \otimes \gras{1}(2,0,\ldots,0)))
 = \frac{d+3}{2d(d+1)},\\
 I^2(1,1,0,\ldots,0) &=&
\frac{1}{\textrm{dim}\mathcal{H}^+_{3}}
\tr( (\gras{1}(1,1,0,\ldots,0) \otimes \proj{\psi_0})( \proj{\psi_0} \otimes \gras{1}(3,0,\ldots,0)))
=\frac{d-1}{d^3+3 d^2+2d},\\
I^2(2,0,\ldots,0) &=& \frac{1}{\textrm{dim}\mathcal{H}^+_{3}} \tr(
(\gras{1}(2,0,\ldots,0) \otimes
\proj{\psi_0})( \proj{\psi_0}
\otimes \gras{1}(3,0,\ldots,0))) =\frac{d+5}{d^3+3 d^2+2d}.
 \eeqa

From these identities, we find can compute the formula
(\ref{eq:varformula}) and obtain the mean variance
(\ref{eq:meanvard}).

\section{The asymptotic limit}

Suppose that one direction is specified by one qubit, and the
other by an infinite number of identically prepared qubits .
We can thus suppose that this second direction, that we choose to call $z$,
is known with arbitrary precision \cite{mass95}. We can therefore imagine that
the first step of our measurement consists in building a \emph{classical} system
that will serve as a $z$-axis. We are thus (again) looking for a povm $\{P(x)\}$
satisfying the conditions
\beq
0 \leq P(x) \leq \gras{1}, \hspace{0.8cm} \int_0^1 dx P(x)=\gras{1}.
\eeq
As a figure of merit, we will consider
\beq
\int dg \int_0^1 dx \bra{\psi_0} \pi(g)^* P(x) \pi(g) \ket{\psi_0} (x-|\bra{\psi_0} \pi(g) \ket{\psi_0}|^2)^2.
\eeq
For any povm $\{P(x)\}$, the povm whose elements are
\bed
Q(x)=\int \frac{d\theta}{2 \pi} e^{-i \theta \sigma_z} P(x) e^{i \theta \sigma_z}
\eed
achieves the same score. We can thus assume that $P(x)$ is diagonal in the $z$-basis:
\bed
\left(
\begin{array}{cc}
s_{0}(x) & 0 \\
0 & s_{1}(x)
\end{array}
\right). \eed The mean error can again be written as \beq
\Delta=\int_0^1 dx (I_0(x) x^2- 2 I_1(x) x+I_2(x)). \eeq Let us
calculate $I_0(x),I_1(x),I_2(x)$. One readily checks that \beqa
I_0(x) &=& \tr P(x) \int dg \pi(g) \ket{\psi_0}\bra{\psi_1} \pi(g)^*, \\
I_1(x) &=& \tr [(P(x) \otimes \proj{\psi_0}) \int dg (\pi^{\otimes 2}(g) \ket{\psi_0^{\otimes 2}}
\bra{\psi_0^{\otimes 2}}
\pi^{\otimes 2}(g)^*
) ], \\
I_2(x) &=& \tr [(P(x) \otimes \proj{\psi_0^{\otimes 2}}) \int dg
(\pi^{\otimes 3}(g) \ket{\psi_0^{\otimes 3}} \bra{\psi_0^{\otimes
3}} \pi^{\otimes 3}(g)^* ) ]. \eeqa Using Shur's lemma, we get
\beqa
I_0(x) &=& \frac{1}{2} \tr P(x)=\frac{1}{2}(s_0(x)+s_1(x)),\\
I_1(x) &=& \frac{1}{3} \tr [ (P(x) \otimes \proj{\psi_0})  S_2]=\frac{1}{3}(s_0(x)+\frac{1}{2}s_1(x)),\\
I_2(x) &=& \frac{1}{4} \tr [ (P(x) \otimes \proj{\psi_0^{\otimes
2}})  S_3]=\frac{1}{4}(s_0(x)+\frac{1}{3}s_1(x)). \eeqa We can
then simply write \beq \Delta=\int_0^1 dx [s_0(x)
(\frac{x^2}{2}-\frac{2}{3}
x+\frac{1}{4})+s_1(x)(\frac{x^2}{2}-\frac{x}{3}+\frac{1}{12})],
\eeq from which we infer that the optimal povm is given by
$s_0(x)=\delta(x-2/3), s_1(x)=\delta(x-1/3)$. In turn the optimal
variance is $\Delta=1/18 \approx .0555$.

\section{The antiparallel case}\label{sec:antipara}

The Clebsch-Gordan series for $\pi^{\otimes 3}$ reads $\pi_{1}^+
\oplus \pi_{1}^+ \oplus \pi_{3}^+$. The problem is now more
complicated because the representation $\pi_1^+$ appears more than
once. As a result, the povm elements do not have an a priori
diagonal form anymore, but only a block-diagonal form (in the
basis corresponding to the irreducible representations): \beq
Q(x)= \left(
\begin{array}{ccc}
q_{00}(x) \gras{1}_{00} &  q_{01}(x) \gras{1}_{01} & \gras{0} \\
q_{10}(x) \gras{1}_{10} &  q_{11}(x) \gras{1}_{11} & \gras{0} \\
 \gras{0} &  \gras{0} & q_{33}(x) \gras{1}_{33} \\
\end{array}
\right). \eeq Explicit expressions for the operators
$\gras{1}_{00},\gras{1}_{01},\gras{1}_{10},\gras{1}_{11},
\gras{1}_{33}$ will be given below.

There exist again functions $I_0(x),I_1(x),I_2(x)$ such that we
can write the average error as $\Delta=\int_0^1 (I_2(x)- 2 I_1(x)
x+ I_0(x) x^2)$. Now, \beq I_i=\int dg \bra{\psi_0,\psi_1,\psi_0}
(\gras{1}^{\otimes 2} \otimes \pi(g)^*) Q(x) (\gras{1}^{\otimes 2}
\otimes \pi(g)) \ket{\psi_0,\psi_1,\psi_0}. |\bra{\psi_0 \pi(g)
\ket{\psi_0}}|^{2i}, \eeq where $i=0,1,2$. We compute these
expressions explicitly. In the following, $\{\ket{0},\ket{1}\}$
will denote an orthonormal basis of the Hilbert space of one
qubit. We start with $I_0(x)$.

\beq
I_0(x)=\tr Q(x) (\proj{01}\otimes \frac{\gras{1}}{2})=
\tr Q(x) ( \int dg \; \pi^{\otimes 2}(g) \proj{01} \pi^{\otimes 2}(g)^*
\otimes \frac{\gras{1}}{2}).
\eeq

Due to Shur's lemma, there exists constants $\gamma_0$ and $\gamma_2$ such
that $\int dg \pi^{\otimes 2}(g) \proj{01}\pi^{\otimes 2}(g)^*=\gamma_0 S_0
+ \gamma_2 S_2$. $S_0$ (resp. $S_2$) is the projector onto the antisymmetric
(resp. symmetric) subspace of two-qubits. Their component (in a computational basis)
are given in terms of Clebsch Gordan-coefficients as
\bed
\bra{uv}S_2 \ket{rs}  \equiv  T_{rs}^{uv}=C^{(2 j)}_{(1 r)(1 s)} C_{(2 j)}^{(1 u)(1 v)},\\
\bra{uv}S_0 \ket{rs}  \equiv  A_{rs}^{uv}=C^{(0 0)}_{(1 r)(1 s)} C_{(0 0)}^{(1 u)(1 v)}.
\eed

(Sum over repeated indices is understood) The constants $\gamma_0$
and $\gamma_2$ are then easily calculated: $\gamma_0= \bra{01} S_0
\ket{01}/ \tr S_0=1/2$, $\gamma_2= \bra{01} S_2 \ket{01}/ \tr
S_2=1/6$. Defining $Q_{rst}^{uvw}(x)=\bra{uvw}Q(x)\ket{rst}$, one
finds that \beq
I_0(x)=\frac{1}{4}Q^{uvw}_{rsw}(x)(\frac{1}{3}T^{rs}_{uv}+A^{rs}_{uv}).
\eeq Similarly, one finds that \beq I_1(x)=\frac{1}{3} \tr (Q(x)
\otimes \proj{0})(\proj{01} \otimes S_2) =\frac{1}{3}
Q_{01t}^{01w}(x) T^{t 0}_{w 0}, \eeq and \beq I_2(x)=\frac{1}{3}
\tr (Q(x) \otimes \proj{00})(\proj{01} \otimes S_3) =\frac{1}{4}
Q_{01t}^{01w}(x) B^{t 0}_{w 0}, \eeq where $B_{rs}^{uv}=C^{(3
j)}_{(1 r)(2 s)} C^{(1 u)(2 v)}_{(3 j)}$.

In the computational basis, the operators $\gras{1}_{ij}$ are explicitly given by

\beqa
(\gras{1}_{00})^{uvw}_{xyz} &=&
C^{(00)}_{(1 x)(1 y)} C^{(1 j)}_{(0 0)(1 z)}C_{(0 0)}^{(1 u)(1/2 v)}C_{(1 j)}^{(0 0)(1 w)},\\
(\gras{1}_{11})^{uvw}_{xyz} &=&
C^{(2 m)}_{(1 x)(1 y)} C^{(1 j)}_{(2 m)(1 z)}C_{(2 k)}^{(1 u)(1 v)}C_{(1 j)}^{(2 k)(1 w)},\\
(\gras{1}_{01})^{uvw}_{xyz} &=&
C^{(2 m)}_{(1 x)(1 y)} C^{(1 j)}_{(2 m)(1 z)}C_{(00)}^{(1 u)(1 v)}C_{(1 j)}^{(0 0)(1 w)},\\
(\gras{1}_{10})^{uvw}_{xyz} &=& [(\gras{1}_{01})^{uvw}_{xyz}]^*,\\
(\gras{1}_{33})^{uvw}_{xyz} &=&
C^{(1 m)}_{(1 x)(1 y)} C^{(3 j)}_{(1 m)(1 z)}C_{(2 k)}^{(1 u)(1 v)} C_{(3 j)}^{(2 k)(1 w)}.\\
\eeqa

With all the information that we have gathered, an explicit calculation can now be carried to find
\beqa
I_0(x) &=& \frac{1}{2} q_{00}(x)+\frac{1}{6} q_{11}(x)+ \frac{1}{3} q_{33}(x),\\
I_1(x) &=& \frac{1}{4} q_{00}(x)+\frac{1}{12}q_{11}(x)
-\frac{1}{12 \sqrt{3}}(q_{01}(x)+q_{10}(x))+\frac{1}{6}q_{33}(x),\\
I_2(x) &=& \frac{1}{6} q_{00}(x)+\frac{1}{18} q_{11}(x)-\frac{1}{12 \sqrt{3}}(q_{01}(x)+q_{10}(x))+\frac{1}{9}q_{33(x)}.\\
\eeqa

As is obvious from the block-diagonal form of $Q(x)$, the error $\Delta$ can be decomposed as $\Delta=\Delta_1+\Delta_3$. $\Delta_3=\int_0^1 dx q_{33}(x) (\frac{1}{9}-\frac{1}{3} h + \frac{1}{3} h^2)$ and $\Delta_1$ can be conveniently written as $\Delta_1= \int_0^1 dx \tr \tilde{Q}(x) F(x)$, where
\bed
\tilde{Q}(x)=
\left(
\begin{array}{cc}
q_{00}(x) &  q_{01}(x)\\
q_{10}(x) &  q_{11}(x)
\end{array}
\right),
\eed
and where

\bed
f_{00}(x) = \frac{1}{6}-\frac{1}{2} x+\frac{1}{2} x^2,\\
f_{11}(x) = \frac{1}{18}-\frac{1}{6} x+\frac{1}{6}x^2,\\
f_{01}(x) = f_{10}(x)= -\frac{1}{6\sqrt{3}} x.
\eed

$\Delta_{33}$ can be readily extremised, setting $q_{33}(x)=\delta(x-1/2)$, giving $\Delta_{33}=1/36$.

The extremisation of $\Delta_1$ is less straightforward. If we
restrict to povm's with a finite number of outcomes, then the
solutions are of the form \beq\label{eq:pesudopauli}
\tilde{Q}^i(x)= w_i^2 \delta(x-x_i) \frac{1}{2} (\gras{1}+ n^i_1
X_1+ n^i_2 X_2+ x^i_3 X_3), \hspace{0.5cm} i=1 \ldots I. \eeq
where $\int_0^1 dx \sum_{i=1}^{\nu} \tilde{Q}^i(x)=\gras{1}$, and
where $X_1,X_2,X_3$ denote the three Pauli matrices.

Then minimising $\Delta_{01}$ amounts to minimise \beq \frac{1}{2}
\sum_{i=1}^{\nu} w_i^2 {f_{00}(x_i)+f_{11}(x_i)+n^i_1
(f_{10}(x_i)+f_{01}(x_i))+n^i_3( f_{00}(x_i)-f_{11}(x_i)}, \eeq
with the constraints $\sum_{i=1}^{\nu} w_i^2=2$, $\sum_{i=1}^{\nu}
w_i^2 n^i_j=0$, $\forall j=1,2,3$. For $\nu=2$, our numerical
extremisation has yielded the following povm:
\beqa\label{eq:optiantipara}
\tilde{Q}^1(x) &=& \frac{1}{2} \delta(x-x_1) (\gras{1}+X_1),\\
\tilde{Q}^2(x) &=& \frac{1}{2} \delta(x-x_2) (\gras{1}-X_1),\\
\eeqa where $x_1=.644338...$ and $x_2=.355662...$. Very
interestingly, the total optimal score $\Delta_{01}+\Delta_{33}$
equals $\Delta^{\textrm{parallel}}(2,1)$. We also wondered whether
increasing the number of outcomes for the part $\Delta_{01}$ could
decrease the overall score. Looking for povms with more outcome,
 we have found no improvement. We therefore believe that
the povm (\ref{eq:optiantipara}) is indeed optimal.

\end{document}